# *Rites de Passage*: Elucidating Displacement to Emplacement of Refugees on Twitter

**Aparup Khatua & Wolfgang Nejdl**

L3S Research Center, Leibniz Universität Hannover, Germany
khatua@l3s.de; nejdl@l3s.de

**Abstract**

Social media deliberations allow to explore refugee-related issues. AI-based studies have investigated refugee issues mostly around a specific event and considered unimodal approaches. Contrarily, we have employed a multimodal architecture for probing the refugee journeys from their home to host nations. We draw insights from Arnold van Gennep's anthropological work '*Les Rites de Passage*', which systematically analyzed an individual's transition from one group or society to another. Based on Gennep's separation-transition-incorporation framework, we have identified four phases of refugee journeys: *Arrival of Refugees*, *Temporal stay at Asylums*, *Rehabilitation,* and *Integration of Refugees* into the host nation. We collected 0.23 million multimodal tweets from April 2020 to March 2021 for testing this proposed framework. We find that a combination of transformer-based language models and state-of-the-art image recognition models, such as fusion of BERT+LSTM and InceptionV4, can outperform unimodal models. Subsequently, to test the practical implication of our proposed model in real-time, we have considered 0.01 million multimodal tweets related to the 2022 Ukrainian refugee crisis. An F1-score of 71.88 % for this 2022 crisis confirms the generalizability of our proposed framework.

## Introduction

*34,361 deaths recorded. Not all the deaths occur at sea, but also in detention blocks, asylum units ... more than 27,000 deaths by drowning since 1993, often hundreds at a time when large ships capsize ... Some entries have a name and a story, but the majority are anonymous data points – just over 1,000 are named ... Some 400 have taken their own lives; more than 600 have died violently at the hands of others.*

*- The Guardian, June 20, 2018*

More than 80 million human beings worldwide have experienced forced displacement by the mid-2020s, and 34 million of them are children below 18 years of age. 26.3 million are refugees, and 4.2 million are stateless people (United Nations Commission on Human Rights Report 2021). The conventional myth is that refugee deaths occur only at sea, but the above excerpt from *The Guardian* counters the same. Refugees face adverse environments not only in the Mediterranean Sea but also after crossing the Mediterranean Sea – if they are lucky. The sufferings of refugees continue at detention blocks or asylums – these sufferings and subsequent deaths are social concerns. Thus, to explore their journey, we draw insights from the seminal work '*Les Rites de Passage*' (1909; The Rites of Passage) by Arnold van Gennep – the French ethnographer and folklorist. Gennep systematically studied the passage of individuals from one social or religious status to another. He has identified three distinct stages of this journey: separation, transition, and re-incorporation. A few studies, mainly from the social science domain, such as Castle and Diarra (2003), Monsutti (2007), have employed this framework in the context of refugees or migrants. However, to the best of our knowledge, none of the prior AI-based studies in the domain of refugee and forced migration studies have employed this framework on social media data to explore the transition of refugees from their home to host nation.

A plethora of studies probed refugee-related issues using microblogging data, such as Twitter (Adler-Nissen et al. 2020; Kreis 2017; Pope & Griffith 2016; Khatua & Nejdl, 2021a; 2021b). While the relevance of multimodal content for extracting relevant and actionable information was widely acknowledged (Alam et al. 2018; Gui et al. 2019), refugee-related studies mostly focused on syntactic text processing (Kreis 2017; Pope & Griffith 2016; Siapera et al. 2018) and did not probe the richness of multimodal data. A handful of studies also explored the emotional responses to refugee-related visual contents (Ibrahim 2018; Olesen 2018). Primarily, the extant literature probed refugee-related deliberations around a specific crisis or event (Pope & Griffith 2016; Siapera et al. 2018). This does not offer a holistic view. BenEzer and Zetter (2015) pointed out that 'the refugee journey is the defining feature of the exilic process: it is a profoundly formative and transformative experience and a 'lens' on the newcomers' social condition'. Thus, we employ Gennep's framework to study their journey.

We have collected 0.23 million multimodal refugee-related tweets during April 2020 to March 2021 and manually annotated 1722 multimodal tweets for the analysis. On the methodology front, the refugee journeys can be conceptualized as a classification task where common subject for our image inputs across all four phases are human beings, i.e., mostly refugees and border forces or NGO staffs – starkly different and challenging from conventional image datasets, such as CIFAR-10, which comprises of distinctly different objects like flowers, animals, and buildings. For our classification task, we have initially considered unimodal models such as Bidirectional Encoder Representations from Transformers (BERT) (Devlin et al. 2019) + Long-Short Term Memory (LSTM) (for text inputs); and InceptionV3 (Szegedy et al. 2016), VGG19 (Simonyan & Zisserman 2015), and ResNet (He et al. 2016) (for visual inputs). Subsequently, we have employed an early fusion of these unimodal models. Figure 1 presents our multimodal framework. Our multimodal models have outperformed unimodal models. The BERT+ LSTM + InceptionV4 (Szegedy et al. 2017) model has reported an accuracy of 80.93%. Subsequently, we collected 0.01 million multimodal tweets related to the 2022 Ukrainian refugee crisis and tested the generalizability of our proposed framework. Our findings, i.e., an F1-score of 71.88 % for this 2022 dataset, strongly indicate that the phases of the refugee journey from displacement to emplacement were identical. Additionally, we have briefly analyzed emotional aspects of our base corpus using LIWC (Tausczik & Pennebaker 2010).

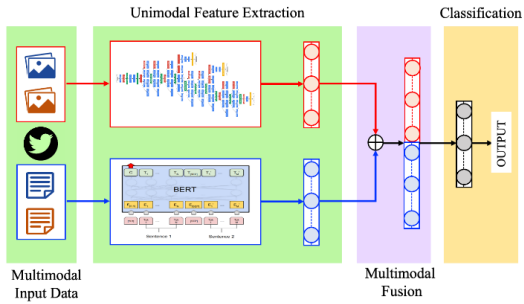

Figure 1: Proposed Multimodal Framework

## Refugee Issues on Social Media Platforms

Refugee journeys can have a wide range of consequences. The significant presence of refugees can impact the economic conditions of the host country (Alloush et al. 2017; Stark 2004). Similarly, a challenging environment in the host country can have adverse consequences on the psychological well-being of refugees (Khawaja et al. 2017). Thus, scholars from multiple disciplines, such as anthropology (Cabot 2016; 2019), economics (Alloush et al. 2017; Stark 2004), management (Karsu et al. 2019), psychology (Brown-Bowers et al. 2015; Khawaja et al. 2017; Maclachlan & McAuliffe 1993), and sociology (FitzGerald & Arar 2018; Karakayali 2018), are probing refugee-related issues. Information science researchers are also investigating voluminous and unstructured social media data, like Twitter deliberations, for understanding various refugee issues. The subsequent sections briefly review this literature.

## Refugee Studies using Text Data

A handful of prior studies analyzed text contents on social media platforms. These studies explored the perception and behavior of social media users mainly from the host nation perspective (Khatua & Nejdl 2022). For example, Siapera et al. (2018) investigated network formation on the Twitter platform in the context of three refugee-related events, namely terrorist attack in Paris, France; sexual assault in Cologne, Germany; and the EU-Turkey deal for refugees. They identified two dominant perspectives - an apprehensive far-right perspective where refugees were considered as terrorists or rapists, and it leads to security and safety concerns in the host countries. Alternatively, a sympathetic and humanitarian view where the discourse revolves around possible ways to help refugees, and some prevalent hashtags were *#safepassage*, *#humanrights*, *#refugeesupport*. Pope & Griffith (2016) have considered a multi-lingual approach and employed sentiment and emotion analysis on 28,866 English and 24,469 German tweets to probe the Twitter deliberations related to Paris and Cologne events. Findings indicate a prevalent presence of negative sentiments - associated with emotions like anger and anxiety. Özerim & Tolay (2020) explored Turkish tweets about Syrian Refugees to understand the effect of echo chamber formation on social media. They found that the Twitter discussions mainly revolved around xenophobic and nationalistic themes. An anti-Syrian hashtag *#ülkemdesuriyeliistemiyorum* (i.e., *#IdontwantSyriansinmycountry* in English) became viral, and it got tweeted and retweeted nearly 50,000 times on a single day in 2016. Similarly, Kreis (2017) analyzed 100 tweets with the hashtag *#refugeesnotwelcome* and found that the discourse of racism against refugees is mainly propounded by the nationalist-conservative and xenophobic right-wing political parties in the European context. Also, they noted hashtags like *#EuropeforEuropeans* by far-right groups in response to the refugee crisis due to the Syrian Civil War. However, in addition to this widespread antipathy and animosity towards migrants and refugees, the literature also noted sympathy and solidarity activities by a certain section of the host nations (Khatua and Nejdl 2022).

Extant literature also probed Twitter text contents for analyzing migration movements across nations. For instance, Urchs et al. (2019) tried to extract locational (to understand 'where the refugees/migrants are headed') and quantitative (to understand 'how many migrants/refugees') information from tweets in the European context. This study has identified 583 tweets about refugees who crossed the border to Hungary, Austria, and Germany in 2015. Similarly, in the context of OECD countries, Zagheni et al. (2014) considered geolocated Twitter data to understand the relationships between internal displacement and international migration.

**Refugee Studies using Image Data**

Images can invoke emotions. Hence, literature probed the emotional responses of mainstream print and social media to the death of the three-year-old Alan Kurdi in the Mediterranean Sea - one of the most unfortunate refugee-related events in our time (Adler-Nissen et al. 2020; Bozdag 2017; Ibrahim 2018; Olesen 2018). The shocking image of the drowned Syrian kid made global headlines on September 2, 2015, and immediately, netizens started sharing it on various social media platforms. This image appeared on the screens of 20 million people worldwide in less than 12 hours after the first release, and it became viral with more than 50,000 tweets per hour (Vis & Goriunova 2015). Adler-Nissen et al. (2020) probed the relationship between images, emotions, and international politics – specifically, how the above image influenced emotional responses, and subsequently, the impact of cumulative emotional responses on international relationships and foreign policy deliberations. They observed that overloaded emotional reactions to the tragic incident of Alan Kurdi had changed the discourse from an open-door approach to stopping refugees from arriving. Similarly, Bozdag (2017) qualitatively analyzed 961 tweets in the context of Turkey and Belgium to investigate the framing and perceptions about refugees before and after the release of Alan Kurdi's image. They did not find any radical shift in the discourse, but the public interest in refugees gained momentum after this tragic incident.

**Why Multimodal Data?**

Multimodal data was widely used in the domain of food items (Wang et al. 2015), e-commerce business (Bi et al. 2017), depression detection (Gui et al. 2019), meme detection (Kiela et al. 2020), and disaster management (Alam et al. 2018) like floods (Ahmad et al. 2019; de Bruijn et al. 2020; Quan et al. 2020). In addition to social media data, prior multimodal studies also considered satellite images in the context of hydrological data (de Bruijn et al. 2020). Multimodal approaches mostly outperform unimodal approaches (Blandfort et al. 2018; Gallo et al. 2020). For instance, hateful meme detection, where opinions are expressed sarcastically, is challenging because it needs to consider both the contextual knowledge and contents of the meme. Here, multimodal approaches are efficient than unimodal approaches (Kiela et al. 2020). Multimodal analysis can perform a wide range of tasks. For instance, a food dataset, such as UPMC Food-101 dataset, which comprises 100,000 recipes for 101 food categories, can be used to analyze and retrieve recipes, dietary assessments, and classification of different food categories (Gallo et al. 2020; Xin Wang et al. 2015). Similarly, in the e-commerce domain, the Rakuten Group has employed multimodal classification to predict each product's type code for defining their catalog (Rychalska & Dąbrowski 2020). Gui et al. (2019) analyzed depression-related tweets. They observed only text contents might fail to detect the depression accurately but considering both text and image contents of a tweet improves accuracy. Multimodal Twitter data was widely used for disaster management. For instance, in the context of natural disasters, such as hurricanes, earthquakes, or wildfire, Alam et al. (2018) argue that retrieving real-time information from the text content and the associated image within a tweet can assist in restoration works. Blandfort et al. (2018) has prepared a multimodal dataset of 1851 tweets to investigate gang violence in the US context. They have identified potential psychological antecedents of violence such as aggression, loss, and substance use, and their multimodal approach outperformed unimodal analysis.

To sum up, prior studies have elucidated the potentials of multimodal approaches for depression detection, disaster management, or identifying psychological antecedents of violent activities. However, refugee-related studies mostly considered unimodal data from social media platforms, i.e., either text or visual content. None of the prior studies explored the potentials of multimodal data for societal transition of refugees from their home to host nation. Thus, we employ a multimodal fusion approach to explore the refugee journeys from displacement to emplacement in the host country (BenEzer and Zetter 2015; Khatua and Nejdl 2021b).

**Refugee Journey:** *Les Rites De Passage?*

*Territorial passages can provide a framework for the discussion of rites of passage … an imaginary line connecting milestones or stakes, is visible – in an exaggerated fashion – only on maps. But not so long ago the passage from one country to another, and, still earlier, even from one manorial domain to another was accompanies by various formalities. These were largely political, legal, and economic, but some were of a magico-religious nature. For instance, Christians, Moslems, and Buddhists were forbidden to enter and stay in portions of the globe which did not adhere to their respective faiths.*

*- Chapter 2*

*… foreigners cannot immediately enter the territory of the tribe or the village; they must prove their intentions from afar and undergo a stage best known in the form of the tedious African palaver. This preliminary stage, whose duration varies, is followed by a transitional period consisting of such events an exchange of gifts, an offer of food by the inhabitants, or the provision of lodging. The ceremony terminates in rites of incorporation – a formal entrance, a meal in common, an exchange of handclasps.*

*- Chapter 3*

Even after 100 years, the above two excerpts from the English translation of the Gennep's antiquarian French work is contemporary and relevant. For instance, there is a striking resemblance between the above century old 'magico-religious' barriers and today's islamophobia in Europe due to middle east refugee-crisis (Zunes 2017). Foreigners (i.e., refugees) cannot immediately enter the territory of the host nations. They need to 'prove their intentions from afar' to border forces, and there's a lot of 'palaver' involved in this process. However, during this period, refugees get food and lodging (i.e., asylums) from inhabitants (i.e., host nations). Finally, rites of incorporation can be equated to the integration of refugees into the society of the host nation. Thus, a handful of prior studies used this theoretical lens in the context of refugee or migration. For instance, Castle and Diarra (2003) note that the migration process of young Malians 'comprises social and psychological dimensions pertaining to the need to explore new places, experience new settings and accumulate material possessions in order to conform to peer group aspirations.' They conclude that the migration of these young Malians 'is as much a rite of passage as a financial necessity' (Castle and Diarra, 2003).

Gennep's research has identified three distinct phases, namely, *pre-liminal rites* (i.e., 'rites of separation from a previous world'), *liminal (or threshold) rites* (i.e., 'those executed during the transitional stage'), and *post-liminal rites* (i.e., 'the ceremonies of incorporation into the new world'). These three stages are commonly referred to as *separation* (getting detached from the previous world and loss of identity), *transition* or liminal stage (the individual has got detached from her previous world and lost her old identity, but not joined the new world), and *incorporation* (getting a new identity after incorporation into the new world). Monsutti (2007) employed this separation-transition-incorporation analogy to explore the journey of young Afghans to Iran. In the initial phase, young Afghans get separated from their families and homes. In the next phase, 'they have to prove their capacity to face hardship and to save money … represents a period of liminality', and finally in the reincorporation phase, they 'return to their village of origin … as adult marriageable men' and mostly they continue to commute between Afghanistan and Iran for the rest of their life (Monsutti 2007). Accordingly, we also employed the Gennep's framework for probing the refugee journeys. In addition to Gennep's work, we have also referred to refugee-related interdisciplinary research to map these three distinct phases of transitions with the refugee journeys. Refugee journeys mostly start because of conflict and violence in their home countries. In many of these countries, freedom of speech is restricted. For instance, in Afghanistan, Facebook allowed its users to lock their profiles instantly and hide friends lists for security concerns. Moreover, internet penetration is significantly low in many of these nations. Social media data does not allow us to analyze this forced displacement process in their home countries. Thus, in this study, refugee journeys start after they plunge into the ocean. If they are lucky, they arrive at their desired destination.

**Phase 1:** We have conceptualized *pre-liminal rites* as the *'Arrival of Refugees'*. Refugees are getting separated from their previous world, getting detached from their families, and taking a leap of faith through risky sea routes as anonymous data points after losing their identity. BenEzer and Zetter (2015) pointed out that the 'mode of travel may influence the meaning of the journey and its impacts on the individual … For someone who has not … crossed the sea before, the mode of travel will be a highly symbolic part of the experience of the journey. These possible differences, and their meaning, need to be investigated'. Thus, for this phase, we have considered tweets deliberating or sharing information about the mode of travel, or tweets related to border control forces.

**Phases 2 & 3:** We have considered two interrelated but distinct stages or activities of *liminal (or threshold) rites* as follows: temporal stay at asylums and rehabilitation of refugees. Our second phase is the *'Temporal stay at Asylums'*. According to Turner (2016), asylums are 'a place of social dissolution and a place of new beginnings where sociality is remoulded in new ways'. Thus, he suggested to 'explore the precarity of life in the camp … in this *temporary space* (emphasis added)'. Accordingly, our tweets in this category capture the details of living conditions in refugee asylums. These tweets also share images of asylums and camps. Our third phase is the *'Rehabilitation of Refugees'*. Khan and Amatya (2017) emphasized the need of health supports because most refugees arrive with health problems ranging from infectious diseases to non-communicable musculoskeletal issues. More importantly, refugees 'face continued disadvantage, poverty and dependence due to lack of cohesive support in their new country … This is compounded by language barriers, impoverishment, and lack of familiarity with the local environment and healthcare system' (Khan and Amatya 2017). Thus, tweets in this category share information and images of various support activities like arrangements of medical aids, donations of food items or garments etc.

**Phase 4:** We have conceptualized the *post-liminal rites* as the *'Integration of Refugees'* into the society of the host nation. Charitable organizations arrange various support activities, such as helping them to learn a new language. For instance, Abou-Khalil et al. (2019) note that Syrian refugees at Lebanon focus on learning English. In contrast, Syrian refugees in Germany try to learn German for better social inclusion. Hence, this category of tweets shares information about the arrangement of the education system for refugee kids or vocational training programs for adult refugees. These skill up-gradation activities help refugees to settle down in the host nations (Bellino & Dryden-Peterson 2019).

Finally, we have also identified one distinct type of multi-modal tweets where text content might be related to the above phases, but image content shares refugee-related statistics or data points through graphical images or charts. These tweets can be crucial for information dissemination in the context of refugee and migrant-related issues. Hence, we have labeled this category of tweets as *'Infographics'*. It is worth noting that this category is not aligned with the Gennep's framework.

# Data

Prior studies, such as Alam et al. (2018), Chen & Dredze (2018), and Gui et al. (2019), have considered the Twitter platform for multimodal analysis. Hence, we have performed keyword-based searching using Twitter's advanced search API, which allowed us to crawl tweets containing a specific keyword. We have considered a set of keywords like *refugee*, *refugee camps*, *refugee asylums*, *migration*, *immigration*, *immigration policy*, etc. We consider English tweets because prior studies on migration observed that the volume of English tweets was significantly higher than other languages (Khatua & Nejdl 2021a). We have collected 3.98 million refugee-related English tweets from April 2020 to March 2021. Our initial analysis indicates that a significant portion of our corpus does not contain images. For our study, we need to consider tweets with images. Subsequently, we also dropped tweets with similar tweet-ids or similar text contents, and our corpus size became 0.23 million multimodal tweets. Our percentage of multimodal tweets (i.e., 5.7% of 3.98 million) is in accordance with prior studies. For instance, Alam et al. (2018) collected 3.5 million tweets during Hurricane Irma, but they found only 0.17 million images i.e., 4.8% multimodal tweets.

**Annotation:** For resource constraints, we selected around 2500 tweets from our corpus of 0.23 million tweets. Our manual annotation process requires fine-grained contextual understanding. For instance, to label the first phase of the journeys, i.e., *'arrival of refugees'*, we have considered the following aspects: arrival through sea routes, risk of traveling through sea routes, mode of transport, activities by border control forces. Similarly, the *'rehabilitation of refugees'* phase has considered the activities such as arranging medical aids, charity activities by non-government organizations (NGOs), or facilitating donations of essential livelihoods. We have carefully analyzed each tweet based on its text content and the associated image for assigning the final class. In this stage, we discarded tweets with poor images or cryptic short texts. We annotated 1722 tweets with distinct text-image pairs from the above sub-sample with an inter-rater reliability of 0.84. These 1722 tweets are distributed as follows: Phase 1: 398 tweets; Phase 2: 387 tweets; Phase 3: 289 tweets; Phase 4: 343 tweets; and Infographics: 305 tweets. Following Madukwe et al. (2020), we have tried to maintain a balance (in terms of tweet volume or percentage) across our five classes. Alam et al. (2018) prepared Twitter-based multimodal datasets for natural disasters or crises like hurricanes, wildfires, earthquakes, and floods. The final volume of annotated tweets for some of their crises were as follows: 1486, 1239, 499, and 832. Hence, our final sample size for analysis is similar to prior Twitter-based studies. We randomly split our 1722 annotated tweets into 80% (as training dataset) and 20% (as test dataset) for our analysis. Table 1 reports a few representative tweets from our corpus.

| | Visual Input | Language Input |
|---|---|---|
| Arrival of Refugees | 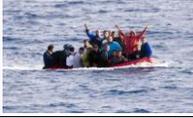 | #Morocco intercepts nearly 200 #migrants trying to reach #Spain: The #Moroccan coast guard intercepted 168 migrants this week who tried to reach Spain using makeshift crafts, including jet-skis and kayaks. |
| | 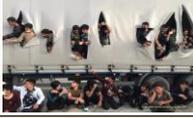 | On August 18, 2020, a total of 167 #migrants were found in the back of a lorry in Samsun, Turkey. The migrants (from Afghanistan and Pakistan) struggled to breathe due to hot weather, and teared the tarpaulin covers during a regular road control by police. |
| Temporal stay at Asylums | 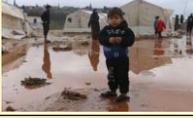 | More than 3,000 Syrians living in Idlib's refugee camps have lost their shelter after days of torrential rain and snow. Although some have left their camps, many families have no choice but to live in flooded tents. |
| | 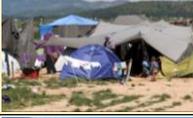 | Proposed EU 'Pact on Migration and Asylum' will not help alleviate migration pressure on EU's southern member states. Nadia Petroni – EU's Pact on Migration and Asylum will do little to ease pressure on southern member states. #refugees #EuropeanUnion |
| Rehabilitation of Refugees | 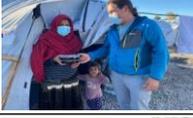 | A recent fire destroyed an entire refugee camp on a Greek island and we were able to respond quickly with food. Our monthly donors allow us to respond nimbly. We couldn't do it without our #CTFriends. Set up your monthly gift here. |
| | 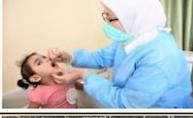 | UNRWA joined a national polio vaccination campaign that started today. Some16000 Palestine refugee children under 5 will be vaccinated at its health centres in Syria. Child immunization is an important part of primary health care @UNRWA provides to Palestine refugees. |
| Integration of Refugees | 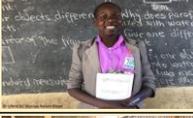 | By purchasing the School Enrollment fees for a refugee girl through our online gift shop, you are helping her survive, recover and build a better future.Help improve her chances of escaping poverty for #InternationalWomensDay |
| | 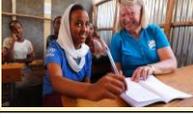 | #DayoftheGirl is so important. Young women and girls still do not have equal access to human rights, education and health. This needs to change. Today I pledge to keep advocating for refugee girls to make sure they can thrive. |
| Infographics | 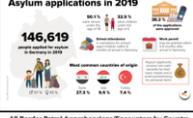 | At the end of 2019, the number of @Refugees worldwide was 79.5 million. #Germany takes in a great amount of refugees, yet the number of #AsylumSeekers has dropped sharply since the refugee crisis of 2015. We give you an insight into the 2019 #asylum statistics. |
| | 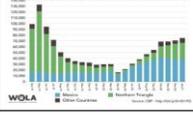 | Here's undocumented migrants apprehended at the US-Mexico border by Border Patrol since April 2019. The largest increase of these 3 categories, by far, from December to January, was "Other Countries"—47%. Arrivals from Mexico and Central America were up only slightly. |

**Table 1:** Representative Multimodal tweets from our Annotated Data (Apr 15, 2020, to Mar 15, 2021)

In contrary to other publicly available image datasets, our image classification is a challenging classification task for computer vision algorithms. Due to the temporal nature of our four phases, common objects across all the categories (except *infographics*) in our corpus are human beings i.e., refugees or migrants. In other words, we have images of refugees arriving through sea routes, or images of refugees with border control forces, or images of refugees at asylums, or images of refugees receiving donation or supports.

## Methodology and Findings

We have initially considered unimodal models for text, i.e., BERT + LSTM, and visual inputs, i.e., InceptionV3 (Szegedy et al. 2016), VGG19 (Simonyan & Zisserman 2015), and ResNet (He et al. 2016). Subsequently, we have also employed an early fusion of the above unimodal models. Deep learning-based models are highly efficient for text classification. For instance, recurrent neural network (RNN) models consider previous information for processing the present computation task. In addition to this basic RNN architecture, LSTM consists of three additional gates: input gate, forget gate, and output gate. LSTM calculates the hidden state by considering the combination of these three gates. In LSTM, the input sequence feed is only in the forward direction. However, a BERT-embedded LSTM can consider tokens from both directions.

Language encoders capture the contextual relation between words either in context-free or by preserving the contextual information. Earlier approaches, such as Word2vec (Mikolov et al. 2013) or Glove (Pennington et al. 2014), generate embeddings for each word without considering their position within a text and surrounding information. However, bi-directional transformer-based models, like BERT, can capture both directions. Thus, BERT models consider the contextual representation of a word to decipher the difference between two contexts (Devlin et al. 2019). Consequently, LSTM models with BERT embedding are more efficient in capturing the contextual representation than LSTM models without BERT embedding (Minaee et al. 2021). We have considered the English language uncased base version of the BERT model with 12 hidden layers, 768 hidden sizes, and 12 self-attention heads.

For image classification, CNN-based computer vision algorithms are most efficient. These models contain convolution layers, max pooling, and fully connected layers to solve complex computer vision tasks. One layer's output becomes the subsequent layer's input in the cascading structure. Similar to language encoders, visual encoders also extract the dominant visual features from an input image and map it into pre-defined categories for the downstream task like image classification, object detection, or instance segmentation. These encoders extract lower dimension features using the deep neural network-based models.

Our first CNN-based model is VGG-19 (Simonyan & Zisserman 2015). VGG takes an image size of $224 \times 224 \times 3$ as an input and performs convolution operation using a $3 \times 3$ filter. We have considered the weights of the VGG19 network based on pre-trained models, but we have trained the output layer using our dataset for the final classification task. Next, we consider ResNet (He et al. 2016) or Residual Networks models. These models skip connection to add the output from a previous layer to a later layer, which tackles the vanishing gradient problem. For ResNet model, we have resized all images to $224 \times 224$ (i.e., rescaled for ease of handling), and the model retains all three RGB channels of the images. We have used ResNet with its ImageNet pre-trained weights to initialize the model, and the outputs are followed by dense layers of 256 units and a SoftMax layer (Deng et al. 2009). Our third CNN architecture-based model is InceptionV3 (Szegedy et al. 2016). This state-of-art third version of Google's initial Inception model explores "ways to scale up net-works in ways that aim at utilizing the added computation as efficiently as possible by suitably factorized convolutions and aggressive regularization" (Szegedy et al. 2016). InceptionV3 is known for its efficiency in interpreting images and detecting objects. We have also considered the fourth version of Inception model for multimodal analysis (Szegedy et al. 2017).

Multimodal fusions leverage the advantages of both modalities to produce a more robust solution for end-to-end system implementation. Thus, multimodal fusion captures the information from content-rich social media data using natural language processing (for textual inputs) and computer vision techniques (for visual contents) for the downstream task. This approach consists of two parallel deep neural architectures. As noted, we employ a transformer-based BERT + LSTM model for textual data and considered multiple pre-trained CNN-based models for visual data. In the final layer, we have employed a fusion-based approach on the outputs from these two models to get the final class from a multimodal tweet (refer to Figure 1).

Our joint multimodal representation can be expressed as $x_m = f(x_1 \cdots x_n)$ where $x_m$ is computed using function $f$ that relies on unimodal representations $x_1 \cdots x_n$. Joint representations are useful for tasks where more than one modality of data is available during training and inference stages. The simplest example of a joint representation is concatenating different modality features at a low level, also known as an early fusion (Baltrusaitis et al. 2019). Early fusion integrates features directly after they are extracted. Early fusion allows to perform multimodal representation learning—as it can learn to exploit the correlation and interactions between low-level features of each modality.

Our fusion considers BERT + LSTM for text inputs and ResNet-50, VGG19, InceptionV3 and InceptionV4 for image inputs. We partially freeze the pre-trained weights and add a dense layer with a dropout layer (p = 0.4), followed by a linear layer to extract the latent features. InceptionV3 and InceptionV4 implementation requires an input size of $299 \times 299$ (i.e., height × width). Hence, we resized the pictures to of $299 \times 299$ for InceptionV3 and InceptionV4. Similarly, we resized the pictures to $224 \times 224$ for VGG19 implementations. We have standardized the pictures using the original ImageNet training mean and standard deviation. Initially, we performed average pooling of $8 \times 8$ for InceptionV3 and InceptionV4 and $7 \times 7$ for VGG. Next, we have applied a dropout of 0.4 and flattened in the next layer. On top of this, we have added a dense unit of 128 before concatenating with the language model stack. In other words, we pass both visual stack and language stack through a shared dense layer of size 128 and concatenate (i.e., early fusion) the outputs to form a joint vector of length 256. We also apply a dense layer of size 256 before the final SoftMax layer and ReLu activation function. This leads to the final classification layer, i.e., a dense layer with 5 units (for our 5 categories) and SoftMax activation, which will give the predicted class of multimodal inputs. Next, we have used stochastic gradient descent (SGD) for model optimization. For robustness, we have used different learning rates for better learning. We train the model using SGD optimizer starting with a learning rate equal to 0.001 and then decreasing it using the Reduce Learning Rate on Plateau from keras which automatically changes the learning rate if there is no improvement in training after a certain number of epochs. We have considered the cross-entropy loss function for our modeling because we have multiple classes in our dataset.

**Findings:** Table 2 reports the accuracies of our unimodal and multimodal models using 5-folds cross-validation. Our BERT + LSTM model for text inputs has reported an F1-score of 58.00%. For brevity, we have not reported the t-SNE plot, but this plot suggests that the weak performance of BERT + LSTM is primarily due to the *Infographics* class, where images differ distinctly from other categories but not the text content. We find VGG19 is our best performing model for the unimodal visual classification task, and the F1-score is 75.30%. F1-scores of our InceptionV3 and ResNet models are 71.52% and 69.35%, respectively. Like prior studies, such as Blandfort et al. (2018), Gallo et al. (2020), we observe that our multimodal models have outperformed unimodal models – except BERT+LSTM + ResNet model. Our BERT + LSTM + InceptionV4 has outperformed other models and reported an accuracy of 80.93%.

| **Models** | **Modality** | **F1-Score** |
|---|---|---|
| *BERT+LSTM (Text)* | | 58.00% |
| *VCG19 (Visual)* | Unimodal | 75.30% |
| *ResNet50 (Visual)* | | 69.35% |
| *Inception V3 (Visual)* | | 71.52% |
| *BERT+LSTM + VGG19* | | 79.69% |
| *BERT+LSTM + ResNet50* | Multimodal | 70.00% |
| *BERT+LSTM + Inception V3* | | 79.06% |
| *BERT+LSTM + InceptionV4* | | **80.93%** |

**Table 2:** Performance of Various Models

**Error Analysis:** Table 3 reports our multimodal models' input text, input image, true label, and predicted label of a few sample tweets. For instance, BERT + LSTM + InceptionV3 has failed to classify the 'Example 4' correctly. The text content of Example 4 indicates their stay in the camp house, but the image is generic – not offering much information about the 4 kids. However, the tweet also talks about the upgradation from 'living in open spaces at the border' to 'private camp house to sleep in'. A significant portion of rehabilitation tweets talks about the upgradation of their status. Probably, our model considers this up-gradation in their living condition as an indicator of 'rehabilitation'.

| | **Example 1** | **Example 2** |
|---|---|---|
| **Input Image** | 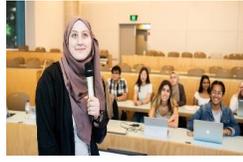 | 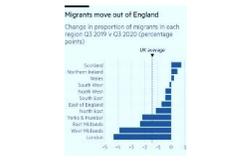 |
| **Input Text** | Help us reach our goal of raising $10,000 (by Dec 10) to support students from refugee and asylum seeker backgrounds at UTS! The funds we raise will help them pay for food, rent, textbooks and cost of living | The importance of immigration and multicultural society. But, Tory Gov supported by Tory Scots ... and their support for English nationalism, Brexit n anti-immigration policies has destroyed Scotland's economy. |
| **True Label** | Integration of Refugees | Infographics |
| **Predicted Label** | Integration of Refugees | Infographics |
| | **Example 3** | **Example 4** |
| **Input Image** | 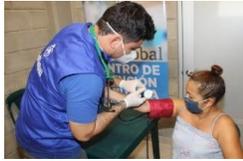 | 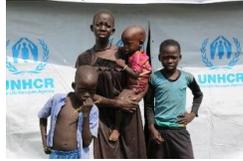 |
| **Input Text** | "It is essential to ensure healthcare during this humanitarian crisis." -Leonardo, nurse on the border, MedGlobal has supported nurses who provide health screenings; first aid, ensuring Venezuelan migrants; | … a refugee mother of 4 children from is happy to have a family shelter in … camp. "My children and I had been living in open spaces at the border for long time. Now I am glad that we finally have our private camp house to sleep in" she said. |
| **True Label** | Rehabilitation of Refugees | Temporal stay at Asylums |
| **Predicted Label** | Rehabilitation of Refugees | Rehabilitation of Refugees |

**Table 3:** Example of sample tweets with classification

## 2022 Ukrainian Refugee Crisis

The previous section leads to the intriguing question: is our proposed framework of four phases generic or context-specific? To probe the practical relevance of our approach, we have considered the 2022 Ukrainian refugee crisis. More than 4 million Ukrainians had to move to neighboring countries, such as Poland, Romania, Hungry, or Moldova, in a span of one and half months. Another 6 million got displaced within the country because of this military invasion (UNHCR Data Portal). Ukrainians got displaced from their previous world and left behind all their belongings (and sometimes family members) towards an uncertain future.

| | Visual Input | Language Input |
|---|---|---|
| Arrival of Refugees | 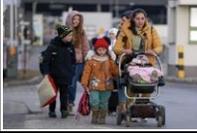 | A Ukrainian refugee pushes her baby in a pushchair as they arrive at the Medyka border crossing, Poland, Saturday, Feb. 26, 2022. |
| | 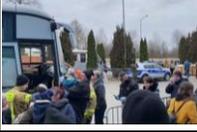 | Every 2 or 3 minutes, a bus full of refugees is arriving into Korczowa's makeshift reception centre. This is what Europe's fastest-growing refugee crisis since WW2 looks like #Poland #Ukraine |
| Temporal stay at Asylums | 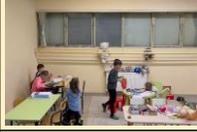 | Our camera is the first inside this refugee center in Rzeszow, #Poland 800 beds, food, clothes.. and this place where kids who've come through bullets and bombs can be kids again. #Ukraine |
| | 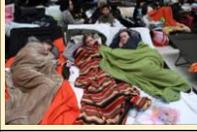 | … People lie on camp beds at a refugee reception centre at the Ukrainian-Polish border crossing in Korczowa, Poland #UkraineRussiaWar #Russia #Ukraine #UkraineWar |
| Rehabilitation of Refugees | 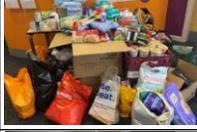 | Great to be in @______ this morning and to see the donations they're collecting to go to Poland, for refugees from #Ukraine Closer to home thanks for all the donations collected for our @____ refugee foodbank too |
| | 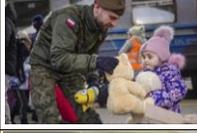 | Polish soldier giving a teddy to a refugee child from Ukraine. #Ukraine #UkraineRussianWar #StandWithUkraine #UkraineRussiaWar #UkraineUnderAttack |
| Integration of Refugees | 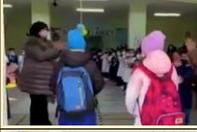 | Ukrainian refugee children fleeing the Russian assault are welcomed with cheers by their Italian peers on their first day at a primary school in Pomigliano d'Arco, Italy, a video shared on social media shows |
| | 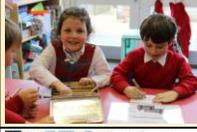 | One of our Durlston Dads is headed to Poland tomorrow with aid for the Ukraine. We have been busy making Happiness Postcards for him to give to the refugee children - a little symbol of love and unity. #DurlstonFamily |
| Infographics | 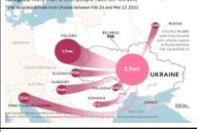 | The number of Ukrainians fleeing the fighting reached 2.7mn by March 13, the UN's refugee agency reported, amid concerns over the growing refugee crisis. The country taking the highest number of refugees is Poland, with 1.7mn alone. |
| | 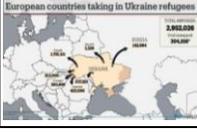 | Nearly one Ukrainian child becomes a refugee every second, as the UN says an average of 73,000 children a day have escaped Russia's onslaught over the last 20 days |

**Table 4:** Representative Multimodal tweets from the 2022 Ukrainian refugee crisis (Feb 24, 2022, to Mar 15, 2022)

**Data:** We have collected 0.6 million tweets, out of which 0.01 million tweets were multimodal, during February 24, 2022, to March 15, 2022. Our data collection and pre-processing strategies were similar to the previous analysis, but for data crawling - we have added a few crisis-specific keywords such as Ukraine, Russia, and so on. Next, we have explored the linguistic and image contents of this corpus. Some of the visible differences are - refugees mostly took the sea routes in the previous corpus, whereas it was mostly rail or road transport during the 2022 crisis. Earlier the travel risk was boat capsizing, and in 2022 it was a missile attack. Similarly, we find images of camps and tents in the previous corpus, whereas it was a temporary makeshift arrangement in hotels and other large public buildings during the 2022 crisis. Broadly, the struggles, risks, and traumas of refugees are identical. A few representative tweets in Table 3 confirm the same. Thus, the intriguing question is: can we apply our proposed framework during this 2022 crisis? If yes, the follow-up question is – whether the previous corpus collected prior to this 2022 crisis (as training data) can help us to extract actionable and relevant information in real-time during the Ukrainian crisis?

**Findings:** To test the same, a single annotator (i.e., one of the authors with prior experience of handling migrant-related tweets) has annotated 234 multimodal tweets (evenly distributed across four phases and the infographics class) and employed our best performing model, i.e., BERT+ LSTM + InceptionV4, on this unseen test data. For a complex unseen dataset like ours, we find that the F1-score came down to 71.88%. As expected, F1-scores of BERT+ LSTM (for unimodal text inputs) and VGG19 (for unimodal visual inputs) models are 50.01% and 59.17%, respectively. Unlike the previous corpus, this 2022 corpus also elucidated the evolution of Twitter deliberations. For instance, initial tweets in February were mainly about the arrival and temporal stays, and in the later period, tweets related to rehabilitation and integration gained momentum.

## Discussion

Refugee-related deliberations have gained momentum in recent times. Twitter gives netizens a platform to raise their voice and share their views. Hence, refugee-related issues are also getting deliberated on Twitter. From the information science domain, prior studies have mostly considered a specific refugee-related event or crisis for their analysis. Opinion mining in the context of one particular event, such as the involvement of a migrant in terrorism in France or unfortunate drowning in the Mediterranean Sea, offers a nuanced understanding of that specific event. However, this approach does not provide a holistic view to activists, refugee workers, and policymakers. To the best of our knowledge, none of the prior studies considered multimodal data to probe refugee journeys from displacement to emplacement.

We have employed the *Les Rites de Passage* framework to elucidate the societal level transitions of refugees from home to host nations. It is worth noting that we have not attempted to track the journey of an individual refugee for ethical concerns but explored the societal level transitions of refugees using multimodal refugee-related tweets - instead of unimodal approach. Our study confirms that this framework can be used to extract relevant and actionable phase-wise information from social media platforms. We note, especially in the context of the 2022 Ukrainian crisis, that refugee needs evolve from essential health support in the initial days to social integration in later days. In the domain of applied AI-based studies, our study may aid policymakers in understanding phase-wise concerns and taking appropriate actions. We conclude that it hardly matters whether someone is taking the risky sea route to reach the shore of the host nation or opting for the rail route or migrant caravan to enter the host nation. The traumatic journey from displacement to emplacement is the same– irrespective of who they are or wherever they come from!

On the methodology front, our research problem is a temporal classification task. Thus, our image inputs, which comprises of mostly human beings (i.e., refugee kids, adult refugees, health workers, border force, educators, and so on) are significantly challenging in comparison to publicly available image datasets like CIFAR-10 that comprises of distinctly different objects like flowers, animals, or buildings. Thus, in the context of applied AI-based research, our study confirms that multimodal models can outperform unimodal models even for a challenging classification task like ours.

**Future Research:** We have also identified a few exciting avenues for future research. For instance, prior NLP-based studies widely used opinion mining of Twitter data. Hence, we have examined the text content of our initial corpus, i.e., 1722 annotated tweets, using linguistic inquiry and word count (LIWC) software (Tausczik & Pennebaker 2010). LIWC is a language tool designed to capture opinions and perceptions by analyzing text contents, and it does not consider the multimodal inputs (which is one potential limitation). Based on our exploratory unimodal analysis, we observe two distinct patterns in our data.

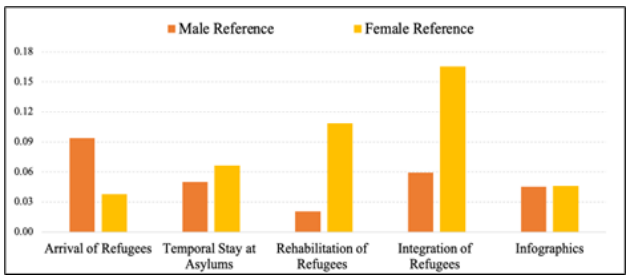

Figure 3: Gender Analysis of our corpus using LIWC

First, we have performed LIWC-based male vis-à-vis female reference analysis, and the vertical axis reports the average score of our 5 categories (refer to Figure 3). We find a distinct pattern. For example, the arrival of refugee class is more associated with male references than female references, whereas female references are significantly higher for rehabilitation and integration classes. Infographics is a gender-neutral class. Probably, our findings indicate that female refugees are getting more supports; whereas safety concerns associated with refugee arrival are more associated with male refugees. Similarly, Rettberg & Gajjala (2016) explored 'the portrayal of male Syrian refugees in a post-9/11 context where the middle eastern male is often primarily cast as a potential terrorist … the claim that the Syrian refugees are primarily male is often repeated on *#refugeesnotwelcome* through images of men with text highlighting the absence of women and children.'

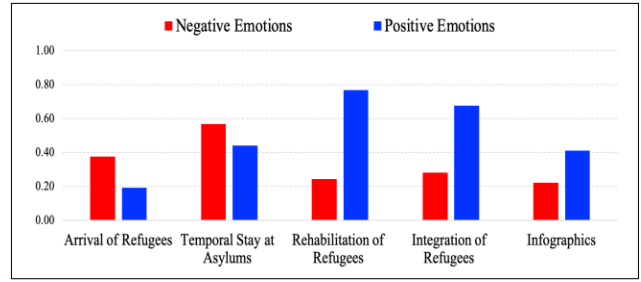

Figure 4: Emotion Analysis of our corpus using LIWC

LIWC-based emotion analysis was also used by prior studies (Saha et al. 2019). It considers 'a number of cognitive strategies, several types of thematic content, and various language composition elements' to capture the positive and negative emotions (Tausczik & Pennebaker 2010). The vertical axis of Figure 4 reports the average score of positive and negative emotions for each category. Once again, we find an interesting pattern. Support activities, such as rehabilitation and integration of refugees, are primarily associated with positive emotions. On the contrary, the first two phases, i.e., arrivals of refugees and temporal stay at asylums, are predominantly displaying negative emotions. The puzzling question is - whether these negative emotions are reflecting xenophobic mindsets? Or social media users are sympathetic to refugee sufferings, and negative emotions reflect the sadness. Future studies need to probe this. Our study elucidates the need to consider these finer nuances of Twitter deliberations for getting an in-depth understanding of refugee-related issues.

# Acknowledgments

Funding for this paper was, in part, provided by the European Union's Horizon 2020 research and innovation program under Grant Agreement No: 832921.